\documentclass[a4paper,fleqn]{cas-dc}

\usepackage[numbers]{natbib}
\usepackage{tabulary}
\usepackage{tabularx}
\usepackage{colortbl}
\usepackage{nccmath}
\usepackage{tikz}
\usepackage{multirow, makecell}
\usepackage[title]{appendix}
\usepackage{framed}

\usepackage{array}
\usepackage{amsmath}
\usetikzlibrary{positioning}
\usepackage{bm}
\usepackage{mathtools}
\usepackage{pbox}
\usepackage{hhline}
\usepackage{url}

\newcolumntype{C}[1]{>{\centering}m{#1}}

\def\tsc#1{\csdef{#1}{\textsc{\lowercase{#1}}\xspace}}
\tsc{WGM}
\tsc{QE}
\tsc{EP}
\tsc{PMS}
\tsc{BEC}
\tsc{DE}

\begin{document}\sloppy
\let\WriteBookmarks\relax
\def\floatpagepagefraction{1}
\def\textpagefraction{.001}
\shorttitle{Community Detection for Access-Control Decisions}
\shortauthors{D\'{i}az Ferreyra et~al.}

\title [mode = title]{Community Detection for Access-Control Decisions: Analysing the Role of Homophily and Information Diffusion in Online Social Networks}           



\author[1]{Nicol\'{a}s E. D\'{i}az~Ferreyra}[orcid=0000-0001-6304-771X]
\cormark[1]
\ead{nicolas.diaz-ferreyra@uni-due.de}
\credit{Conceptualization, Methodology, Investigation, Writing- Original Draft Preparation, Formal Analysis, Data Curation}

\author[2]{Tobias Hecking}[orcid=0000-0003-0833-7989]
\ead{tobias.hecking@dlr.de}
\credit{Methodology, Investigation, Validation, Writing- Reviewing and Editing}

\author[3]{Esma A\"{i}meur}
\ead{aimeur@iro.umontreal.ca}
\credit{Validation, Writing- Reviewing and Editing}

\author[1]{Maritta Heisel}
\ead{maritta.heisel@uni-due.de}
\credit{Validation, Writing- Reviewing and Editing}

\author[1]{H. Ulrich Hoppe}
\ead{hoppe@collide.info}
\credit{Validation, Writing- Reviewing and Editing}

\address[1]{University of Duisburg-Essen, Department of Computer Science and Applied Cognitive Science, 47057 Duisburg, Germany}
\address[2]{Institute for Software Technology Intelligent and Distributed Systems, German Aerospace Center, 51147 K\"{o}ln, Germany}
\address[3]{Department of Computer Science and Operations Research, University of Montr\'{e}al, Canada}

\cortext[cor1]{Corresponding author}


\begin{abstract}
Access-Control Lists (ACLs) (a.k.a. ``friend lists'') are one of the most important privacy features of Online Social Networks (OSNs) as they allow users to restrict the audience of their publications. Nevertheless, creating and maintaining custom ACLs can introduce a high cognitive burden on average OSNs users since it normally requires assessing the trustworthiness of a large number of contacts. In principle, community detection algorithms can be leveraged to support the generation of ACLs by mapping a set of examples (i.e. contacts labelled as ``untrusted'') to the emerging communities inside the user's ego-network. However, unlike users' access-control preferences, traditional community-detection algorithms do not take the \textit{homophily} characteristics of such communities into account (i.e. attributes shared among members). Consequently, this strategy may lead to inaccurate ACL configurations and privacy breaches under certain homophily scenarios. This work investigates the use of community-detection algorithms for the automatic generation of ACLs in OSNs. Particularly, it analyses the performance of the aforementioned approach under different homophily conditions through a simulation model. Furthermore, since private information may reach the scope of untrusted recipients through the re-sharing affordances of OSNs, information diffusion processes are also modelled and taken explicitly into account. Altogether, the removal of gatekeeper nodes is further explored as a strategy to counteract unwanted data dissemination.
\end{abstract}

%



\maketitle

\section{Introduction} \label{intro}

Online Social Networks (OSNs) like Twitter or Facebook are virtual spaces that allow people to connect with like-minded others by exchanging different types of media content including posts, links, and pictures \cite{penni2017future}. To a large extent, social interaction inside these platforms resembles communication aspects of everyday life. Particularly, the exchange of private information, either inside or outside OSNs, is fundamental for creating and maintaining social relationships \cite{wang2016modeling}. Hence, it is not surprising that people often disclose personal information in social media platforms in order to strengthen their bonds and maximize their social capital \cite{kramer2019mastering}. Nevertheless, keeping private information away from untrusted recipients becomes a challenging task for average users since OSNs place the members of disjoint social circles (e.g., family and work colleagues) under a same communication channel \cite{kramer2019mastering, vitak2012impact}. Consequently, this often leads to unintentional privacy breaches due to misalignments between the \textit{intended} and the \textit{actual} audience of online publications \cite{schwartz2020selectivity}.

Access-Control Lists (ACLs) or just ``friend lists'' are one of the most salient privacy mechanisms of OSNs since they allow users to constrain the audience of the content they publish online \cite{hirschprung2017analyzing,ni2016empirical}. Basically, ACLs are collections of contacts that are deemed \textit{untrusted} by the user regarding the access to certain pieces of personal information. In principle, ACLs are an effective way to keep the information disclosed inside posts away from unintended recipients. For example, an ACL composed of work colleagues could be applied to restrict the visibility of a post with a negative comment about one's employer. However, creating custom ACLs can introduce a high cognitive burden on average users since it demands assessing the trustworthiness of a large number of contacts \cite{namara2018potential,ahmed2019information}. Furthermore, different ACLs must be created for different types of personal information since individuals' trustworthiness may vary from content to content \cite{dong2016ppm}. Last but not least, keeping ACLs' internal consistency can demand a great effort given that users are likely to add or remove contacts from their network over time \cite{mondal2019moving}.

\begin{figure}[pos=!b]
\centering
\includegraphics[width=\linewidth]{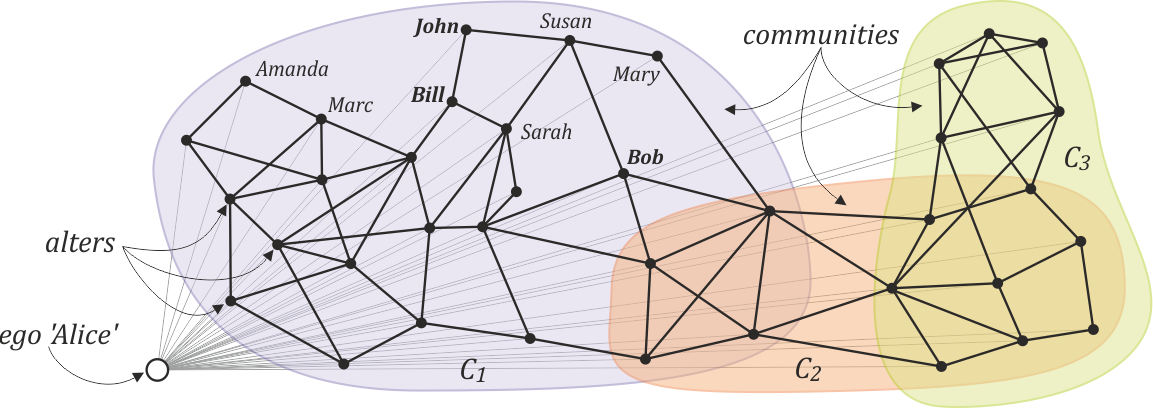}
\caption{Alice's ego-network. Untrusted friends John, Bill, and Bob are grouped under cluster $C_1$. An ACL containing all the nodes in $C_1$ is generated and recommended to Alice \cite{diaz2018access}.}
\label{fig:ego_example}
\end{figure}

Access-Control Predictive Models (ACPMs) aim to reduce the burden of manual configurations through the automatic generation of custom ACLs \cite{ahmed2019information}. Particularly, ACPMs leverage a set of \textit{classification features} (e.g. personal attributes or network structure) to elaborate and recommend ACLs aligned with a set of \textit{access-control preferences} provided by the user \cite{khazaei2016detecting,ni2016empirical}. One approach consists of applying community detection algorithms for identifying clusters of untrusted members inside the user's ego- network (i.e. the network of connections between her friends) \cite{misra2016non, khazaei2016detecting}. Under this approach, an ACL is created out of the members of the cluster that best fits the user's access-control preferences. In this case, preferences are specified through a set of contacts marked as ``untrusted'' by the user (Fig.~\ref{fig:ego_example}). Such a community-based ACPM is suitable particularly in cases where accessing the personal attributes of network members (e.g. age, gender, workplace) is limited or not possible at all \cite{misra2017react,misra2016non}. This is because traditional community-detection algorithms (such as Leading Eigenvector \cite{newman2006finding} or Multilevel Community \cite{blondel2008fast}) can identify clusters without requiring information on the attribute values of its nodes (if any).

\subsection*{Motivation}

Although community-based ACPMs only require structural network information for their application, they can nonetheless lead to inaccurate ACL configurations. Particularly, this is due to the influence of \textit{homophily} in the formation of communities inside a network as on people's access-control preferences \cite{towhidi2020trust,de2018discovering,gao2014friends}. Basically, homophily refers to an organization principle in which the members of a network with similar characteristics tend to create connections with each other \cite{mcpherson2001birds}. On the one hand, such a principle has a large impact on a network's structure and consequently in the formation of its communities \cite{de2018discovering}. Particularly, it is to expect that the members of an emerging cluster would hold a certain similarity with regard to a set of attribute values (e.g. location, gender or workplace) \cite{kim2017effect}. On the other hand, users' access-control preferences are also largely influenced by some of the attribute values shared among network members. For instance, in order to generate an ACL of work colleagues, a user would label as ``untrusted'' some of the contacts who share her same ``workplace'' attribute value. However, unlike the user's ACL preferences, traditional community-detection algorithms are not driven by attribute similarity causing flaws in the identification of untrusted communities \cite{diaz2018access,misra2016non,gao2013circles}. In particular, they may select best-fit clusters whose members do not portray similarities on the attribute values deemed relevant for the generation of the corresponding ACL \cite{diaz2018access}. Consequently, such a community-based ACPM may generate inaccurate ACL configurations and privacy vulnerabilities under certain homophily scenarios.

Another factor that can affect the efficacy of ACLs is information diffusion in OSNs. Particularly, OSNs are endowed with affordances that allow their users to re-share content from others making it available to an audience beyond the one defined by the content owner \cite{humbert2019survey}. This can result in privacy breaches especially when a \textit{trusted} member makes the user's post available to an \textit{untrusted} one after re-sharing it \cite{humbert2019survey,yu2018my}. Nevertheless, despite its importance in terms of privacy and access-control, the dynamics of information diffusion in OSNs are often neglected by ACPMs including community-based ones \cite{backes2017reconciling}. Consequently, privacy violations can occur even when ACLs manage to properly identify all the untrusted recipients of a particular piece of private information. This calls for the elaboration of more robust solutions that are not only capable of satisfying users' access-control preferences, but also manage to prevent the propagation of personal information throughout the untrusted segments of a network.

\subsection*{Contribution}

This work elaborates on the findings reported in \cite{diaz2018access} further investigating the impact of homophily and information diffusion in community-based ACPMs. Particularly, the performance of this approach is evaluated against different network topologies and information diffusion settings. For this, two simulation experiments are outlined, executed, and interpreted. In the first one, topologies are generated according to particular homophily conditions describing the attachment probability between nodes of similar characteristics. These network configurations are used thereafter to determine the precision, recall, and F1 score of community-detection algorithms when predicting ACLs. Unlike in \cite{diaz2018access}, Links in Context \cite{hecking2018links}, an attribute-based clustering approach, is also considered for performance evaluation along with traditional structure-based methods. In the second experiment, the dynamics of information diffusion (not considered in \cite{diaz2018access}) is introduced in the generated networks to elaborate countermeasures against unwanted data dissemination. In this case, the removal of gatekeeper nodes (i.e. trusted members linked to untrusted clusters) is investigated through a model of social influence as an approach for reducing the amount of personal information spread across untrusted segments of a network.

The rest of this paper is organized as follows. In the next section, related work on access-control prediction and information diffusion in OSNs is discussed and analysed. Following, Section \ref{background} introduces the theoretical foundations of this paper. In particular, homophily-driven preferential attachment and social influence models are presented and described for later application. Sections \ref{experiment1} and \ref{experiment2} describe the methodology and results of simulation experiments 1 and 2, respectively. Such results together with their limitations are further discussed in Section \ref{discussion}. Finally, in Section \ref{conclusions}, we outline the conclusions of this paper and introduce directions for future work.

\section{Related Work} \label{related_work}

Controlling the access to personal information is until now one of the main privacy challenges of OSNs. Particularly, such a challenge has motivated several contributions dealing with the automatic configuration and recommendation of ACLs. This section discusses related work elaborating on different ACPMs within the current literature. Likewise, advances on information diffusion and social influence are also analysed under the lens of access-control decisions.

\subsection{Access-Control Prediction} \label{rw_acpm}

Privacy and security scholars have developed several approaches to facilitate the definition of access-control policies in OSNs. Particularly, ACPMs seek to automate the generation of ACLs through machine learning \cite{ni2016empirical,dong2016ppm,misra2017pacman,diaz2018pst,shan2020smart}, formal logic \cite{rafiq2017learning,vahabli2019novel}, and network analysis \cite{gao2013circles,misra2017react,diaz2018access} among other methods. On a large scale, ACPMs can be classified into \textit{community-based} \cite{gao2013circles,misra2017react,diaz2018access} or \textit{attribute-based} \cite{ni2016empirical,dong2016ppm,diaz2018pst,squicciarini2014identifying}, depending on whether they leverage communities or personal attributes for the automatic generation of access-control policies. For instance, D\'{i}az Ferreyra et al. \cite{diaz2018pst} introduced an attribute-based solution in which decision trees are generated to recommend adequate post audiences in OSNs. Under this approach, friends are classified into trusted or untrusted by applying a number of conditional tests over a set of profile attributes such as age, gender, interest and education. A similar strategy is followed by Dong et al. \cite{dong2016ppm} who elaborated on a classifier in which the sharing tendency between users together with the sensitivity of the content being disclosed are employed as audience predictors. Similar attributes along with demographic and location-related information were used by Ni et al. \cite{ni2016empirical} in a machine learning solution that recommends personalized privacy policies for user-generated content in OSNs.

Despite their levels of accuracy, attribute-based solutions display some limitations related to the information on which access-control predictions are made. Particularly, deciding which attributes should be deemed as predictors is not a trivial decision. Furthermore, some attributes may not be available across different OSNs, making it difficult to engineer multi-platform solutions \cite{misra2017react}. Conversely, community-based ACPMs do not suffer from these limitations and may be considered as less privacy invasive since their predictions are grounded on clustering algorithms that use solely the network structure as input. For example, Misra et al. \cite{misra2017react} unveiled untrusted social circles in OSNs using the Clique Percolation Method, an approach which builds up communities out of fully-connected network sub-graphs. Certainly, hybrid solutions have also been introduced and discussed along with community and attribute-based ACPMs. Such is the case of Fangfang et al. \cite{shan2020smart} who elaborated a model for the automatic trust assessment of network members based on both, the connections and attribute similarities among them. Nevertheless, although hybrid solutions can show a high accuracy, their performance tends to decrease as the size of the network becomes larger \cite{ding2019novel}. Furthermore, a performance comparable to the one of hybrid ACPMs can be also achieved by community-based solutions within a shorter time frame \cite{misra2017react,gao2013circles}.

\subsection{Information Diffusion in OSNs} \label{rw_diffusion}

A large body of research has been dedicated to the study of diffusion processes in complex networks \cite{dhamal2018effectiveness,purba2020influence,li2018forecasting, jiablocking2020,backes2017reconciling,gorla2019enhanced}. Particularly, a significant number of works have explored approaches for the identification of the most influential users in a network with the aim of \textit{maximizing} the spread of information \cite{li2018influence}. For instance, Dhamal \cite{dhamal2018effectiveness} elaborated on influence optimization through an adaptive seed strategy consisting of multiple phases of information diffusion. Under this method, influence maximization is achieved by selecting seed nodes at different stages according to the observed propagation of information over time. In line with this, Radion Purba et al. \cite{purba2020influence} introduced an information diffusion model that incorporates the engagement and activeness levels of Instagram users as indicators of their influence susceptibility and influence degree, respectively. Work concerning the identification of diffusion participants can also be found within the current literature. Such is the case of Li et al. \cite{li2018forecasting} who developed a method for spotting members that are likely to forward viral information in Twitter. Particularly, they elaborated a ranking of potential hashtag adopters by analysing and forecasting the use of hashtags across multiple chains of followers. 

All in all, influence maximization research is of great value for many real-world applications including online marketing \cite{zeng2020business}, trending topic detection \cite{miao2016cost}, and information summarization \cite{shi2015vegas}. Nevertheless, information diffusion has also been analysed from an influence \textit{minimization} perspective and applied in areas such as public health policies \cite{zhang2019data}, misinformation \cite{yan2019rumor}, and cybersecurity \cite{jiablocking2020,backes2017reconciling,gorla2019enhanced}. For instance, Jia et al. \cite{jiablocking2020} proposed limiting the propagation of adversarial content through the deletion of critical nodes and edges in a network. Likewise, Yan et al. \cite{yan2019rumor} employed an edge-removal strategy to counteract the diffusion of rumours across OSNs. Overall, these applications share the common goal of stopping the spread of an agent (viruses, information, etc.) across the members of a network. However, access-control decisions in OSNs introduce the additional challenge of optimizing the utility of information disclosure. That is, maximizing the number of trusted recipients while reducing the number of untrusted ones. Prior work has introduced formal frameworks for analysing this problem and algorithms for approximating its solution (cf. \cite{backes2017reconciling,gorla2019enhanced}). Still, to the best of our knowledge, such a challenge has not been tackled from a community-detection perspective. Furthermore, it has not been analysed yet under the lens of homophily. 

\section{Theoretical Background} \label{background}

As discussed in Section \ref{rw_diffusion}, ACPMs often neglect the role of information diffusion when generating custom privacy policies. Moreover, homophily can also affect the performance of community-based approaches since it has a direct impact in the formation of communities inside OSNs. This work aims to investigate these aspects through simulation models of information diffusion and attributed scale-free networks. Such models and their main characteristics are introduced and discussed in the following subsections.

\subsection{Homophilic Preferential Attachment} \label{background_attachment}

Large complex networks like the Internet or OSNs have caught the attention of researchers across different fields \cite{barabasi2009scale}. Empirical studies have shown that these networks share an important property: only a small number of nodes hold a big amount of connections to other nodes, whereas most nodes have just a few \cite{barabasi2016network}. Networks containing such important nodes, or \textit{hubs}, tend to be \textit{scale-free} in the sense that the degree of these hubs (i.e. number of connections to other nodes) widely exceeds the average \cite{jiang2017structure,barabasi2016network}. Up to now, scholars have proposed several evolution models for constructing scale-free networks \cite{khan2018online,kim2017effect,dangalchev2004generation,barabasi1999emergence}. Among those models, one of the most prominent ones is the one introduced by Barabasi and Albert \cite{barabasi1999emergence}. This model indicates that two simple mechanisms, \textit{growth} and \textit{preferential attachment}, are responsible for the emergence of scale-free networks \cite{barabasi2016network}. On the one hand, growth refers to the process in which at each time step a node with $m$ ($\leq m_0$) links is added to the network and connected to $m$ pre-existing nodes. Preferential attachment, on the other hand, describes the process by which new nodes prefer to link to the more connected nodes in the network (i.e. the hubs) \cite{barabasi2016network}. 

In principle, the preferential attachment mechanism proposed by Barabasi and Albert does not take attribute similarity into consideration. Nevertheless, this approach has been enriched with homophily characteristics by other researchers resulting in more specific network evolution models \cite{khan2018online,kim2017effect,diaz2018access}. For example, Kim et al. \cite{kim2017effect} introduced a \textit{group-openness} mechanism for modelling the homophily and attachment probability between two nodes in a network. Under this approach, a node characterised with the attribute value $s$ is considered a member of the group $s$. Then, the attachment probability between a node $i$ of group $s$ and a node $j$ of group $t$ is computed as a function of the openness factor $\Lambda_{s}^{t}$ between the groups $s$ and $t$. Such a factor can adopt values between 0 and 1 and indicates how closed (or open) is a node in group $t$ to create links with other nodes in $s$ \cite{kim210evolution}. Diaz Ferreyra et al. \cite{diaz2018access} adopted and extended this mechanism to situations in which nodes are characterized through multiple attribute values and therefore deemed members of more than one group. Particularly, homophily across different groups is expressed through an openness matrix $\boldsymbol{\Lambda}$ composed by the openness factors of all pairs of attribute values available in the network (see Appendix \ref{model} for a detailed description). The corresponding preferential attachment model is adopted in this work for the generation of network topologies aligned with specific homophily conditions.

\subsection{Independent Cascade} \label{ic_model}

Many efforts have been made to understand and recreate information diffusion processes in OSNs \cite{li2017survey,chang2018study,li2018social}. To a wide extent, such efforts have their origins in studies seeking to simulate the dynamics of epidemic diseases in biological networks. Overall, this is because information (just like a virus) begins to spread from a set of seed nodes to the rest of the network at a certain diffusion rate \cite{chang2018study,li2018social}. Different information diffusion models within the current literature have sought to recreate such a process making particular assumptions about its dynamics \cite{chang2018study}. For example, progressive models like the Lineal Threshold (LT) \cite{shakarian2015independent} assume that, once infected, nodes cannot switch back to their original non-infected state. Conversely, non-progressive models like the Susceptible Infected Susceptible (SIS) \cite{newman2003structure} consider the scenario in which an infected node can return to its initial condition and therefore be infected many times. Particularly a progressive variation of this last one, the Susceptible Infected Recovered (SIR), has been applied in the context of the COVID-19 pandemic to simulate chains of contagion across communities \cite{cooper2020sir}.

Another approach widely applied to describe information diffusion processes in OSNs is the Independent Cascade (IC) \cite{li2018social,shakarian2015independent}. Let us consider an ego-network representation $G=(V,E)$ where the node set $V$ corresponds to the user's befriended contacts and $E$ to the connections existing between them. The IC is a non-progressive model in which information spreads from an initial set of infected nodes $A_{0} \subseteq V $ to the rest of the network members like a domino \cite{chang2018study}. For the case of sharing a post in an OSN, $A_{0}$ could be defined as a sub-group of friends who notice the user's post once published and are likely to re-share it. Hence, at each time step $t \geq 0$ an infected node $v \in A_{t}$ can pass the information to an inactive (i.e. non-infected) neighbour $u : (u,v) \in E$ with probability $p_{vu}$. If successful, $u$ becomes infected at time step $t+1$, otherwise $u$ stays inactive and $v$ has no chance to infect it again. However, if $u$ has more than one infected neighbour, it will be approached by each of them independently. Such a contagion process continues to unfold until there are no more infection attempts to be triggered. In Section \ref{exp2_method}, the IC model is adopted and instantiated to evaluate i) the effectiveness of the predicted ACLs under information diffusion conditions, and ii) elaborate countermeasures to prevent unwanted data dissemination in OSNs.

\section{Simulation 1: Generation of ACLs} \label{experiment1}

All in all, homophily is a process that may impact a network's structure and consequently impair the outcome of community-based ACPMs. This experiment aims to analyse the effect of homophily when generating ACLs from the emerging communities inside ego-networks. For this, such a community-based approach was put into practice in simulated network topologies under different homophily conditions. The following subsections describe the methodology employed in the experiment and the results obtained from its execution.

\subsection{Methodology}

\begin{figure*}[pos=!t, align=\centering]
\includegraphics[height=5.1cm]{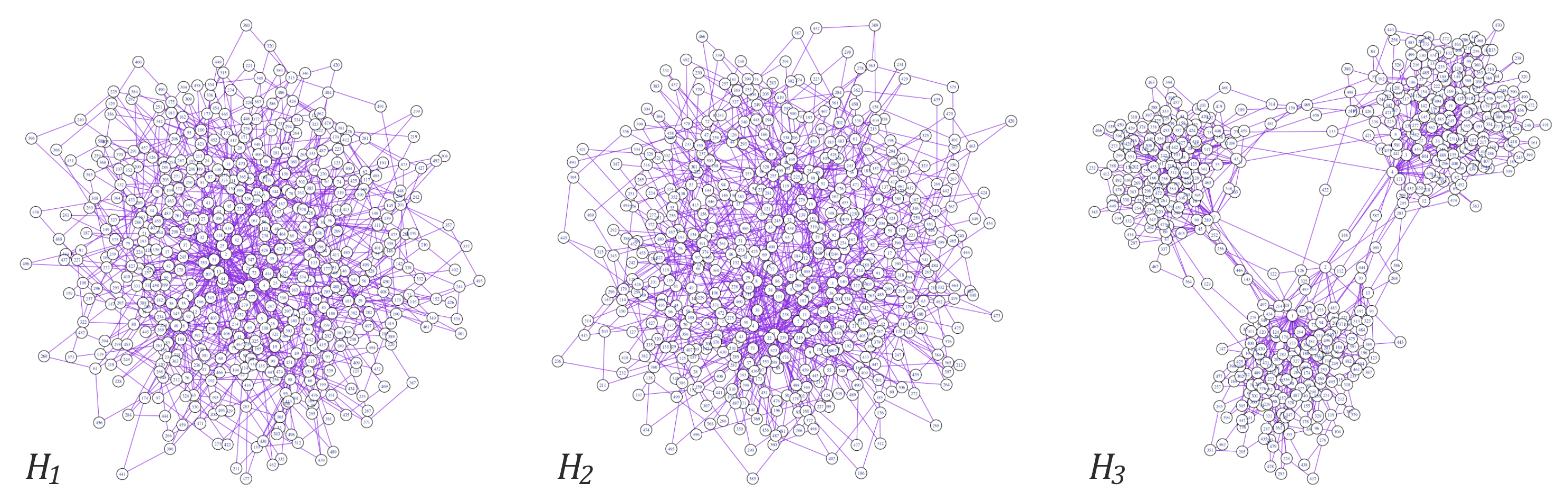}
\caption{Simulation output for homophily conditions $H_1$, $H_2$, and $H_3$.}
\label{fig:networks}
\end{figure*}

In this experiment, the simulation approach introduced in \cite{diaz2018access} is followed. As it is shown in Fig.~\ref{fig:pipeline1}, a community-based ACPM requires (i) the user's ego-network and (ii) her privacy preferences in order to generate a personalized ACL. Particularly, such preferences can be described through a small sample of untrusted friends. That is, some of those contacts who should be excluded from the audience of a certain piece of personal information. Therefore, this stage consists of a method for simulating ego-networks and a criterion for the selection of untrusted network members:

\begin{figure}[pos=!b]
\centering
\includegraphics[width=\linewidth]{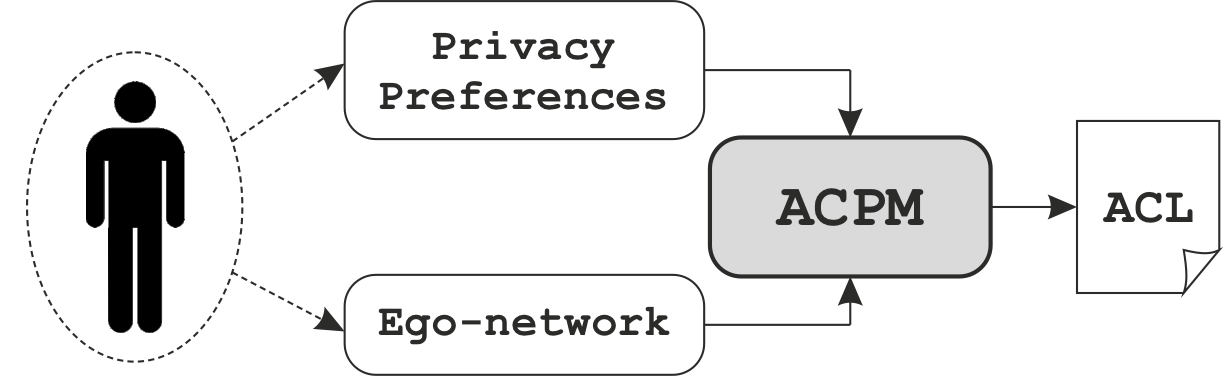}
\caption{Simulation approach for the generation of ACLs \cite{diaz2018access}.}
\label{fig:pipeline1}
\end{figure}

\begin{itemize}
\item \textit{Simulation of ego-networks}: As discussed in Section \ref{background_attachment}, networks are generated through a homophily-driven preferential attachment model in which nodes are characterized with the attributes \textit{gender}, \textit{workplace}, and \textit{location}. Particularly, nodes in the generated networks can adopt the values $male$ or $female$ for the gender attribute, $Starbucks$, $Google$, or $Ikea$ for workplace, and $Leeds$, or $York$ as location. Under this model, the attachment probability between two nodes is computed through an openness matrix $\boldsymbol{\Lambda}$ describing the homophily conditions of the network. Basically, the values inside $\boldsymbol{\Lambda}$ can range from 0 to 1 and represent the strength of attribute similarity in the linking process (e.g. a value $\Lambda_{York}^{Leeds}$ closer to zero describes a setting in which users located in \textit{York} are less likely to connect with users living in \textit{Leeds}). An extended description of this mechanism can be found in the Appendix.

\item \textit{Selection of untrusted nodes}: The selection of untrusted network members is guided by a hypothetical self-disclosure scenario in which a user working in $Ikea$ (i.e. the \textit{ego} of the simulated network) wishes to create an ACL to exclude her work colleagues from the audience of her publications. Hence, $n$ nodes with the attribute value $workplace=Ikea$ are selected from the generated ego-networks representing the user's access-control preferences. Particularly, nodes with the highest degree are selected as these are often the most influential ones in the network. The corresponding ACL is then built out of the community that brings together the largest amount of these untrusted nodes. For this, a community-detection algorithm (CDA) is employed to identify clusters of nodes inside the network under analysis.
\end{itemize}

The goal of this simulation experiment is to evaluate the performance of community-based ACLs under different homophily conditions. For this, a set of network topologies were generated out of different $\boldsymbol{\Lambda}$ values and used thereafter to produce the corresponding ACLs. Since $workplace$ is the attribute that drives the selection of untrusted nodes, three $\boldsymbol{\Lambda}$ configurations were defined to simulate networks with a higher/lower degree of $workplace$ homophily:
\begin{itemize} \setlength\itemsep{1ex}
\item $H_1:\{\Lambda_{Google}^{Starbucks}=\Lambda_{Starbucks}^{Ikea}=\Lambda_{Google}^{Ikea}=0.7\}$
\item $H_2:\{\Lambda_{Google}^{Starbucks}=\Lambda_{Starbucks}^{Ikea}=\Lambda_{Google}^{Ikea}=0.3\}$
\item $H_3:\{\Lambda_{Google}^{Starbucks}=\Lambda_{Starbucks}^{Ikea}=\Lambda_{Google}^{Ikea}=0.01\}$
\end{itemize}
As it can be observed, these configurations differ only in the values assigned to $\Lambda_{Google}^{Starbucks}$, $\Lambda_{Starbucks}^{Ikea}$ and $\Lambda_{Google}^{Ikea}$ while the rest of the group-openness factors were set to one. These values were adopted from \cite{diaz2018access} as they lead to topologies with significant differences in structure and workplace homophily.

\subsection{Execution and Results} \label{exp1_results}

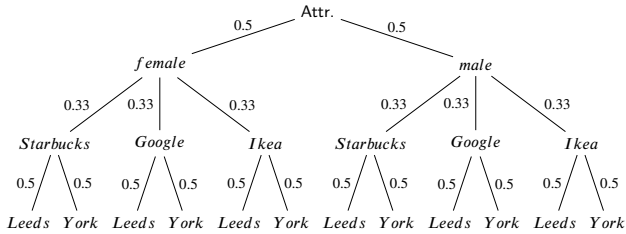
\begin{figure}[pos=!h]
\centering
\resizebox{\linewidth}{!}{
\begin{tikzpicture}[
level 1/.style={sibling distance=6cm, level distance = 1cm},
level 2/.style={sibling distance=2cm, level distance = 1.5cm},
level 3/.style={sibling distance=1cm, level distance = 1.5cm},
arc/.style={inner sep=3mm, font=\footnotesize},
grow'=down,
scale=0.95, 
every node/.style={scale=0.95}]
    \node (Root) [] {Attr.}
        child [black] {
        node {$male$}
        child { node {$Ikea$} 
                child {
                		node[] {$York$}
                    	edge from parent
                		node[right, font=\footnotesize] {$0.5$}}
                child {
                		node[] {$Leeds$}
                   	edge from parent
                		node[left, font=\footnotesize] {$0.5$}}
                edge from parent
                node[right, arc] {$0.33$}
        }
        child [black] { node {$Google$} 
                child {
                		node[] {$York$}
                    	edge from parent
                		node[right, font=\footnotesize] {$0.5$}}
                child {
                		node[] {$Leeds$}
                   	edge from parent
                		node[left, font=\footnotesize] {$0.5$}}
                edge from parent
                node[left, font = \footnotesize] {$0.33$}
        }
        child [black] { node {$Starbucks$} 
                child {
                		node[] {$York$}
                    	edge from parent
                		node[right, font=\footnotesize] {$0.5$}}
                child {
                		node[] {$Leeds$}
                   	edge from parent
                		node[left, font=\footnotesize] {$0.5$}}
                edge from parent
                node[left, arc] {$0.33$}
        }
        edge from parent
        node[above, font=\footnotesize] {$0.5$}
    }
    child {
        node {$female$}
        child { node {$Ikea$} 
                child {
                		node[] {$York$}
                    	edge from parent
                		node[right, font=\footnotesize] {$0.5$}}
                child {
                		node[] {$Leeds$}
                   	edge from parent
                		node[left, font=\footnotesize] {$0.5$}}
                edge from parent
                node[right, arc] {$0.33$}
        }
        child [black] { node {$Google$} 
                child {
                		node[] {$York$}
                    	edge from parent
                		node[right, font=\footnotesize] {$0.5$}}
                child {
                		node[] {$Leeds$}
                   	edge from parent
                		node[left, font=\footnotesize] {$0.5$}}
                edge from parent
                node[left, font=\footnotesize] {$0.33$}
        }
        child [black] { node {$Starbucks$} 
                child {
                		node[] {$York$}
                    	edge from parent
                		node[right, font=\footnotesize] {$0.5$}}
                child {
                		node[] {$Leeds$}
                   	edge from parent
                		node[left, font=\footnotesize] {$0.5$}}
                edge from parent
                node[left, arc] {$0.33$}
        }
        edge from parent
        node[above, font=\footnotesize] {$0.5$}
    };
\end{tikzpicture}}
\caption{Attributes probability distribution \cite{diaz2018access}.}
\label{fig:prob_tree}
\end{figure}

The simulation model described in the previous subsection was implemented using iGraph \cite{igraph2006}, a library for network analysis and visualization for R\footnote{Link to the code repository: \url{https://bit.ly/35kGmqZ}}. A total of three ego-networks of size $N=500$ were simulated each aligned with the homophily conditions $H_1$, $H_2$ and $H_3$. In all cases, the attribute values of the nodes were assigned following the probability tree of Fig. \ref{fig:prob_tree}. Likewise, $n=10$ untrusted nodes were selected from each of the networks for the generation of the corresponding ACLs. Particularly, three different CDA were applied to unveil clusters inside the simulated networks: Leading Eigenvector (LE) \cite{newman2006finding}, Multilevel Community (MC) \cite{blondel2008fast}, and Links in Context (LiC) \cite{hecking2018links}. Both LE and MC were initially assessed in a prior study (c.f. \cite{diaz2018access}). These methods rely solely on structural information to split the nodes of the network into a hierarchy of nested communities. Conversely, LiC also leverage the network's attributes to identify densely-connected groups of nodes \cite{hecking2018links}. Basically, LiC defines the context of each link in the network as the subset of attributes that are common to its endpoints. Thereby, it identifies communities of nodes through agglomerations of links sharing the same context. To the best of our knowledge, LiC has not been yet evaluated for the automatic generation of ACLs.

\begin{table}[pos=b!]
\centering
\bgroup
\def\arraystretch{1.2}
\resizebox{\linewidth}{!}{
\begin{tabular}{|C{0.8cm}|C{0.8cm}|C{1.2cm}|C{0.6cm}|C{1.2cm}|C{1cm}|C{1cm}|}
\hline
\multicolumn{1}{|C{0.8cm}|}{\textbf{Conf.}} & 
\multicolumn{1}{C{0.8cm}|}{\textbf{CDA}} & 
\multicolumn{1}{C{1.2cm}|}{\textbf{N\textsuperscript{o} of Clusters}} & 
\multicolumn{1}{C{0.6cm}|}{\textbf{ACL Size}} & 
\multicolumn{1}{C{1.2cm}|}{\textbf{Precision}} & 
\multicolumn{1}{C{1cm}|}{\textbf{Recall}} & 
\multicolumn{1}{C{1cm}|}{\textbf{F1}} \tabularnewline \hline
 \multirow{3}{*}{$H_1$} & MC & $15$ &$53$ & $30.19\%$ & $8.94\%$ & $13.79\%$ \tabularnewline \hhline{|~|-|-|-|-|-|-|}
  & LE & $28$ &$49$ & $48.98\%$ & $13.41\%$ & $21.05\%$ \tabularnewline
\hhline{|~|-|-|-|-|-|-|}
  & LiC & $234$ &$272$ & $36.03\%$ & $54.75\%$ & $43.46\%$ \tabularnewline \hline
 \multirow{3}{*}{$H_2$} & MC & $15$ &$67$ & $50.75\%$ & $21.25\%$ & $29.96\%$ \tabularnewline \hhline{|~|-|-|-|-|-|-|}
  & LE & $16$ &$40$ & $75.00\%$ & $18.75\%$ & $30.00\%$ \tabularnewline 
\hhline{|~|-|-|-|-|-|-|}
  & LiC & $530$ &$53$ & $62.26\%$ & $26.63\%$ & $30.99\%$ \tabularnewline \hline
 \multirow{3}{*}{$H_3$} & MC & $6$ &$141$ & $99.29\%$ & $81.87\%$ & $89.74\%$ \tabularnewline \hhline{|~|-|-|-|-|-|-|}
  & LE & $6$ &$172$ & $97.67\%$ & $98.25\%$ & $97.96\%$ \tabularnewline 
\hhline{|~|-|-|-|-|-|-|}
  & LiC & $32$ &$175$ & $97.14\%$ & $99.42\%$ & $98.27\%$ \tabularnewline \hline
\end{tabular}
}
\egroup
\caption{Simulation 1: Results of configurations $H_1$, $H_2$, and $H_3$.}
\label{table:sim1_results}
\end{table}

Fig.~\ref{fig:networks} illustrates the network topologies generated by the simulations. One can observe that, as the workplace openness becomes smaller (i.e. small $\Lambda_{Google}^{Starbucks}$, $\Lambda_{Starbucks}^{Ikea}$ and $\Lambda_{Google}^{Ikea}$), the nodes of the network tend to agglomerate around three big communities. Hence, the precision of the predicted ACLs (i.e. the percentage of untrusted nodes it contains) improves considerably from $H_1$ to $H_2$ and reaches its maximum in the homophily condition $H_3$ (Table~\ref{table:sim1_results}). Particularly, the highest precision in $H_1$ and $H_2$ is achieved through the LE algorithm with a $48.98\%$ and $75.00\%$, respectively. In the case of $H_3$ the best precision score corresponds to the algorithm MC with a value of $99.29\%$. On the other hand, the method's recall (i.e. percentage of untrusted nodes in the network included in the generated ACL) also reaches its maximum value in $H_3$ for all the clustering methods. However, only MC and LE improve their respective recall values from configurations $H_1$ to $H_2$. Conversely, the recall of LiC drops from $54.7\%$ in $H_1$ to $26.63\%$ in $H_2$. The same phenomenon is observed for the F1 score (i.e the weighted average of precision and recall) which in the case of LiC decreases from $43.46\%$ in $H_1$ to $30.99\%$ in $H_2$. Nonetheless, LiC obtained the highest recall in $H_1$ and $H_3$ and the best F1 score across all homophily configurations.

In terms of execution time, MC and LE were able to identify clusters within seconds whereas LiC took more than 3 minutes in the best case ($H_3$)\footnote{Values obtained with a 1.3 GHz dual-core processor and 8 GB 1867 MHz memory.}. With respect to the number of generated clusters, LiC unveiled more communities than LE and MC in all configurations. Particularly, LiC identified the largest number of clusters in $H_2$ (530 clusters) whereas the smaller amount corresponds to MC and LE in $H_3$ (6 clusters each). As for the ACL size, the average number of clustered members was $124.67\pm127.61$ in $H_1$, $53.33\pm13.50$ in $H_2$, and $162.67\pm18.82$ for the homophily condition $H_3$. 

\section{Simulation 2: Information Diffusion} \label{experiment2}

As discussed in Section \ref{rw_diffusion}, the re-sharing affordances of OSNs can impair the effectiveness of ACLs and violate, in turn, the privacy preferences of content-owners. This section introduces a simulation experiment for analysing the performance of community-based ACLs under different information diffusion settings. For this, the network topologies generated in Section \ref{experiment1} are infected using the IC model introduced in Section \ref{ic_model}. Particularly, an ANOVA test is conducted to determine the effects of the network's homophily and the number of seed nodes on the percentage of infected nodes at the corresponding ACL. Furthermore, the effects of removing gatekeeper nodes are also evaluated and proposed as a means for counteracting unwanted data dissemination.

\subsection{Methodology} \label{exp2_method}

\begin{figure}[pos=!b]
\centering
\includegraphics[width=\linewidth]{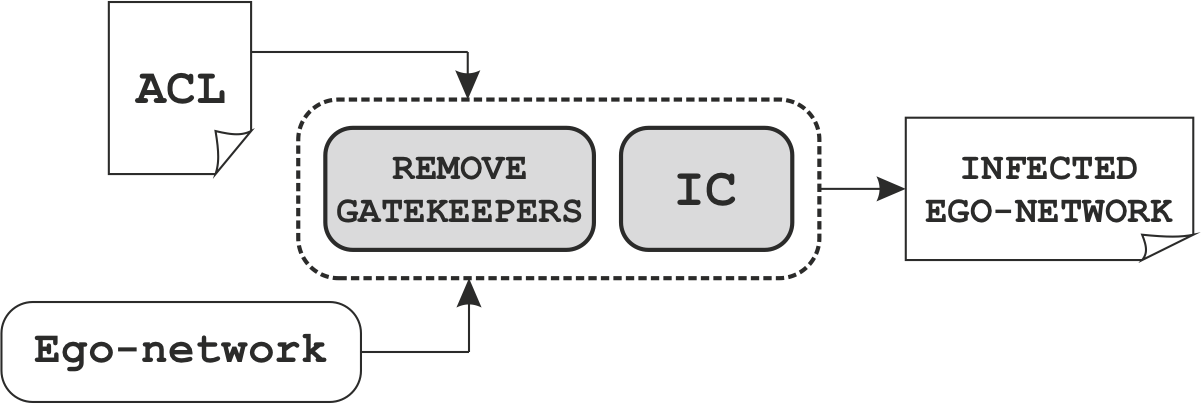}
\caption{Simulation approach for information diffusion.}
\label{fig:pipeline2}
\end{figure}

Fig.~\ref{fig:pipeline2} illustrates the main building blocks of the proposed simulation approach. First, a total of $k$ gatekeeper nodes of the highest degree are removed from the ego-network under analysis. Such nodes are trusted members linked to one or more nodes inside the corresponding ACL and may therefore forward the information to untrusted network segments. Next, a group of $s$ seed nodes is selected from outside the ACL and infected according to the premises of the IC model. Basically, this step recreates the situation in which the user's post reaches members of the trusted audience that are likely to re-share it. In principle, the IC model assumes that all infected nodes in the network can pass the information to their non-infected neighbours with the same probability. However, to a certain extent, the dynamics of information diffusion are related to homophily characteristics of OSNs \cite{DeChoudhury2010}. Particularly, the flow of user-generated content across OSNs can be strongly influenced by attribute similarities among its users \cite{qin2020homophily}. Therefore, the IC model was adjusted so that the probability with which an infected node $v$ spreads information to a non-infected neighbour $u$ is computed as:

\begin{ceqn}
\begin{align*}
p_{vu}=& \frac{\#SharedAttVal}{\#TotalAtt} \cdot \beta
\end{align*}
\end{ceqn}

where $\#TotalAtt$ corresponds to the number of attributes characterizing the network (3 in this case) and $\#SharedAttVal$ to the number of attribute values shared between $v$ and $u$. The parameter $\beta$ represents the maximum infection probability among all network members and can adopt values ranging between 0 and 1.

\subsection{Execution and Results}

\begin{table*}[pos=!t]
\def\arraystretch{1.3}
\centering
\caption{Factorial ANOVA test.}
\begin{tabular}{ l c c c c c c } 
\hline 
\multicolumn{1}{c}{\textbf{Source}} & 
\textbf{Sum Sq.} &
\textbf{d.f.} & 
\textbf{Mean Sq.} &
\textbf{F} & 
\textbf{\emph{p}} & 
\textbf{$\eta^2$} \tabularnewline
\hline 

Homophily & 17050.551 & 2 & 8525.276 & 77.622 & 0.000 & 0.080\tabularnewline
  
Seeds & 7988.157 & 1 & 7988.157 & 72.731 & 0.000 & 0.039 \tabularnewline

Rm-Gatekeepers & 686064.725 & 2 & 343032.363 & 3123.266 & 0.000 & 0.778 \tabularnewline

Homophily * Seeds & 1562.111 & 2 & 781.056 & 7.111 & 0.001 & 0.008 \tabularnewline

Homophily * Rm-Gatekeepers & 43990.027 & 4 & 10997.507 & 100.131 & 0.000 & 0.184 \tabularnewline

Rm-Gatekeepers * Seeds & 8761.669 & 2 & 438.335 & 3.991 & 0.019 & 0.004\tabularnewline

Homophily * Seeds * Rm-Gatekeepers & 884.828 & 4 & 221.207 & 2.014 & 0.090 & 0.005 \tabularnewline

\hline
\end{tabular}\\ 
\begin{tabular}{@{}c@{}} 
\end{tabular}
\label{table:sim2_results}
\end{table*}

\begin{table*}[pos=!t]
\def\arraystretch{1.3}
\centering
\caption{Tukey HSD Post-Hoc Test}
\begin{tabular}{ c c c c } 
\hline 
\multicolumn{1}{c}{\textbf{Pair}} & 
\textbf{Mean Diff.} &
\textbf{\emph{p}} &
\textbf{95\% CI} \tabularnewline
\hline 
$H_2$ - $H_1$ & 6\% & 0.000 & (4.58\%, 7.42\%) \tabularnewline
$H_3$ - $H_1$ & -0.96\% & 0.253 & (-2.38\%, 0.46\%) \tabularnewline
$H_3$ - $H_2$ & -6.96\% & 0.000 & (-8.37\%, -5.54\%) \tabularnewline
$\frac{1}{3}$ Removed - $0$ Removed & -30.95\% & 0.000 & (-32.37\%, -29.53\%) \tabularnewline
$\frac{2}{3}$ Removed - $0$ Removed & -47.05\% & 0.000 & (-48.46\%, -45.63\%) \tabularnewline
$\frac{2}{3}$ Removed - $\frac{1}{3}$ Removed & -16.09\% & 0.000 & (-17.51\%, -14.67\%) \vspace{0.5ex}\tabularnewline
\hline
\end{tabular}\\ 
\begin{tabular}{@{}c@{}} 
\end{tabular}
\label{table:tukey}
\end{table*}

The effects of homophily along with the number of seed nodes and removed gatekeeper nodes were analysed in a 3x2x3 factorial experimental design. Particularly, $k=$ 0, $\frac{1}{3}$ or $\frac{2}{3}$ of the getaway nodes were removed from the networks corresponding to the homophily configurations $H_1$, $H_2$ and $H_3$. In addition, these networks were infected with $s=$ 75 or 150 seed nodes as described in Section \ref{exp2_method}. In all cases, the $\beta$ parameter was set to 0.6 to represent a tendency of the network members towards information re-sharing. Likewise, all ACLs were generated using the LE algorithm since it was the one with the best overall precision in Simulation 1. The IC model was executed 100 times for each of the 3x2x3=18 experimental conditions resulting in 1800 simulation runs. 

An ANOVA test was conducted on the simulation results to determine the effects of i) homophily, ii) diffusion seeds, and iii) removed gatekeepers on the percentage of infected ACL members. At first, a Levene's test determined that the dataset resulting from the simulation outputs did not meet ANOVA's assumption of homogeneity of variance. However, this condition can be relaxed for large samples (N $\geq 30$) and when the number of observations is equally distributed (or nearly so) across the different factors and levels \cite{sawyer2009analysis}. Since these conditions are met by the generated data set (i.e. N=1800 and 100 observations per level), the ANOVA analysis was carried out accordingly. As summarized in Table~\ref{table:sim2_results}, all of the main factors were significant for the percentage of infected ACL members after the execution of the IC model ($p>0.001$). Particularly \emph{homophily} yielded an effect size of $\eta^2 = 0.080$, indicating that around 8\% of the variance in the percentage of infected ACL nodes was explained by this factor ($F_{2,1782}=77.622, p=0.000$). Such a variance is affected in 3.9\% by the \emph{diffusion seeds} ($F_{1,1782}=72.731, p=0.000$) and in 77.8\% by the \emph{removed gatekeepers} ($F_{2,1782}=3123.266, p=0.000$). A significant interaction effect of 0.008 was observed between \emph{homophily} and \emph{diffusion seeds} ($F_{2,1782}=7.111, p=0.001$). Likewise, a significant effect of 0.184 was yielded from the interaction between \emph{homophily} and \emph{removed gatekeepers} ($F_{4,1782}=100.131, p=0.000$). However, no significant effects were observed from the interaction of all three factors altogether ($F_{4,1782}=2.014, p=0.090$).

In order to further examine the significant main effects, a Tukey HSD Post-Hoc test was conducted (Table~\ref{table:tukey}). Particularly, an average drop of 30.95\% in the infected ACL nodes was observed after removing $\frac{1}{3}$ of the gatekeeper nodes and 47.05\% after removing $\frac{2}{3}$ ($p<0.001$). Likewise, the average difference of -16.10\% between the $\frac{2}{3}$ and $\frac{1}{3}$ gatekeeper removal conditions was also found significant ($p<0.001$). On the other hand, the percentage of infected ACL members increases 6\% on average from homophily condition $H_1$ to $H_2$ and decreases about 6.96\% from $H_2$ to $H_3$ ($p<0.001$). However, no significant differences were observed between homophily conditions $H_3$ and $H_1$.

\begin{figure*}[pos=!t, align=\centering]
\includegraphics[height=5.1cm]{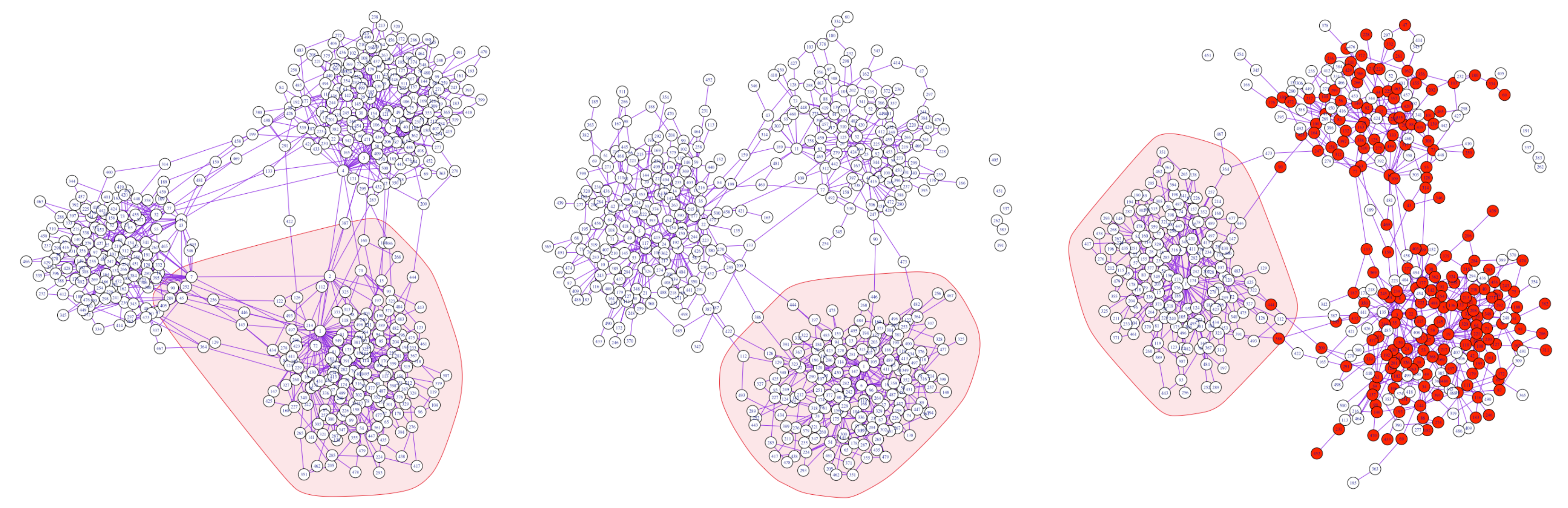}
\caption{Infection sequence for $H_3$: ACL (left), after removing 2/3 of gatekeepers (center), after information diffusion with 150 seeds (right).}
\label{fig:h3_sequence}
\end{figure*}

\section{Discussion} \label{discussion}

Several remarks can be drawn upon the results obtained from the simulation experiments. On the one hand, the performance of community-based ACPMs is closely related to the homophily of the network under analysis and the attribute values that are critical for the ACL configuration. This can be observed from the results of Simulation 1 where the precision of the predicted ACL becomes higher as the homophily in workplace increases. Particularly, LE resulted the method with the highest overall precision making it more suitable than MC and LiC for cases in which the costs of false positives (i.e. classifying trusted members as untrusted) are deemed high. Conversely, the recall and F1 values obtained in this experiment suggest that LiC would qualify better for cases where the penalty of false negatives (i.e. classifying untrusted members as trusted) is larger. In part, this is explained by the number and size of the communities generated by this method which, compared to MC and LE, result much larger. Nonetheless, the recall and F1 scores of LE are close to the ones of LiC in both, $H_2$ and $H_3$, making it also a good approach in such a case with the advantage of being computationally more effective.



Overall, the F1 values of LiC show that its loss in terms of precision is not so large. However, this metric along with the method's recall did not improve from configuration $H_1$ to $H_2$ (something that did happen with MC and LE). Furthermore, LiC produced much more false positives than MC or LE making it less suitable for cases in which Type 1 errors are considered critical. In part, this may be related to the approach followed by this method when partitioning the network space into overlapping communities. Particularly, LiC determines the optimal cut level of the resulting dendrogram through a measure of its partition density which tends to divide the network into many small communities. Moreover, such communities may in turn be included in larger ones which is not the case for LE nor MC. This has a direct impact in the detection of the community that best fits the user's privacy preferences in terms of size and composition. Hence, such an aspect should be further investigated and adjusted so the LiC algorithm becomes more suitable for access-control prediction.


On the other hand, the results of Simulation 2 suggest that homophily can impact significantly the performance of community-based ACLs under information diffusion conditions. Likewise, such a performance is also affected by the number of seed nodes spreading the information across the network at $t=0$. Nevertheless, the largest effect size and the highest drop in the percentage of infected ACL members is given by the number of gatekeeper nodes removed from the network. Certainly, it is to expect that as more gatekeeper nodes get removed, fewer infections inside the corresponding ACL will be observed. However, this has also a negative impact on the trusted audience's extent since gatekeepers are, after all, nothing but trusted network members. Furthermore, as observed in Fig.~\ref{fig:h3_sequence} some trusted nodes get disconnected from the network as a consequence of applying such a removal strategy. Hence, such a collateral damage should be minimized so the influence range of the content being disclosed is not significantly reduced. Particularly, these side effects should be taken explicitly into consideration for the elaboration of an adequate gatekeeper removal criterion. 

To a certain extent, the results yielded by the simulation experiments of this work are subject to limitations. Particularly, such results may be affected by the set-up parameters of the simulation models as well as by the size and attribute characterization of the generated networks. On the one hand, the average size of ego-networks can largely differ from one OSN to another. For instance, it was estimated that around 38.35\% of Facebook users in the United States had between 500 and 200 friends by 2016\footnote{statista: Average number of Facebook friends of users in the United States in 2016 - \url{https://bit.ly/3oljuyY}}. However, this number can easily reach the order of thousands in other OSNs like Instagram\footnote{statista: Worldwide Instagram follower growth rate from January to June 2019, by profile size - \url{https://bit.ly/2JPlxwo}}. On the other hand, the type of attributes proposed in Section \ref{exp1_results} as well as the distribution of their values may not necessarily represent actual OSNs conditions. Instead, it is to expect that certain attribute values will prevail over others and may not be equally distributed across the network members \cite{volkovich2012length}. Finally, the probability of influence among OSNs users is subject to factors that go beyond attribute similarity. For instance, it can be affected by individual relations, time, and network centrality \cite{li2018influence}. Hence, further experimental studies should closely consider these factors in order to characterize the IC model more adequately.

\section{Conclusion and Future Work} \label{conclusions}

The experimental results of this work provide valuable insights into the role of homophily and information diffusion when applying community-based ACPMs. Particularly homophily was shown to be a critical aspect for the generation of adequate ACL configurations through community-detection methods. In this sense, it is of paramount importance to have a good understanding of the homophily conditions at the network under analysis before applying clustering methods for ACL prediction. On the other hand, avoiding unwanted data dissemination remains an open challenge for access-control policies in OSNs. In principle, this issue can be mitigated by excluding a percentage of gatekeeper nodes from the members of the trusted audience. Nevertheless, side effects on the utility of information disclosure should be taken into consideration when applying such a countermeasure. For this, a removal criterion of gatekeeper nodes should seek to maximize the influence of the shared information while minimizing the chances of its propagation towards untrusted network segments.

There are multiple questions and future research directions that arise from the results of this work. One of them corresponds to conducting a study of community-based ACPMs under actual network conditions. This includes an analysis of attribute similarity and influence processes in OSNs for an adequate characterization of the simulation models. Likewise, evaluating other community detection approaches for access-control prediction will be a matter of future investigations. On the other hand, we expect to explore the models and principles presented in this work on areas outside privacy and security. Particularly the removal of gatekeeper nodes under certain homophily conditions could be applied to the design of public health policies, for instance, in elaborating social distancing strategies to control the spread of the COVID-19 pandemic. In such a case, more specific models for the simulation of epidemic diseases will be investigated and evaluated along with the principles of gatekeeper removal discussed throughout this work.

\printcredits

\section*{Acknowledgements}
This work was partially supported by the H2020 European Project No. 787034 ``PDP4E: Privacy and Data Protection Methods for Engineering'' and Canada's Natural Sciences and Engineering Research Council (NSERC).

\vskip3pt

\bibliographystyle{cas-model2-names}

\bibliography{references_difussion}

\clearpage
\begin{appendices}
\section{Group Openness} \label{model}

As introduced in section \ref{background_attachment}, Kim et al. \cite{kim2017effect} defined a \textit{group-openness} mechanism to investigate the role of homophily in the evolution of scale-free networks. Particularly, such approach considers that nodes sharing a particular attribute value $s$ are members of a common group $s$. Then, the group-openness factor $\Lambda_s^t$ between two groups $s$ and $t$ is defined as:

\begin{align} \label{eq1}
\Lambda_{s}^{t}=
\begin{cases}
    \Lambda,& \text{if } s\neq t\\
    1,              & \text{if } s=t
\end{cases}  \parbox{2.5cm}{\footnotesize\raggedleft $0 \leq \Lambda \leq 1$}
\end{align}

where the homophily index $\Lambda$ is a real number between 0 and 1 \cite{kim210evolution}. Particularly a value of $\Lambda = 0$ describes the case in which nodes in $s$ are completely reluctant to create ties with others who are members of group $t$. Conversely, for $\Lambda = 1$, members of $s$ connect openly with other nodes outside their group. Such $\Lambda$ values can be employed to define how strong (or weak) is the role of homophily between two nodes in terms of group membership. 

\newcommand\scalemath[2]{\scalebox{#1}{\mbox{\ensuremath{\displaystyle #2}}}}

\begin{figure}[pos=!h]\centering
\[\scalemath{1.00}{
\renewcommand{\arraystretch}{2}
\bm{\Lambda}
=
\begin{bmatrix}
    \Lambda_{male}^{male} & \Lambda_{male}^{female} &  \dots  & \Lambda_{male}^{York} \\
    \Lambda_{female}^{male} & \Lambda_{female}^{female} & \dots  & \Lambda_{female}^{York} \\
    \vdots & \vdots & \ddots & \vdots \\
    \Lambda_{York}^{male} & \Lambda_{York}^{female} & \dots  & \Lambda_{York}^{York}
\end{bmatrix}
=
\begin{bmatrix*}[c]
    1~ & 0.7~ & \dots~  & 0.3 \\
    0.7~ & 1~ & \dots~  & 0.5 \\
    \vdots & \vdots & \ddots~ & \vdots \\
    0.3~ & 0.5~ & \dots~  & 1
\end{bmatrix*}}
\]
\caption{Group-openness Matrix.}
\label{fig:matrix}
\end{figure}

As mentioned in Section \ref{experiment1}, the nodes of the networks analysed in this work are characterized with the attributes gender, workplace and location. Particularly, a node can adopt the values \textit{male} or \textit{female} for the ``gender'' attribute, \textit{Starbucks}, \textit{Google}, or \textit{Ikea} for ``workplace'', and \textit{York} or \textit{Leeds} for ``location''. Therefore, each network consists of 7 groups making it possible to define up to $C_{7,2}=\frac{7!}{2!(7-2)!}=21$ group openness factors. For instance, $\Lambda_{York}^{Leeds}$ can be used to specify how open (or closed) is people from $York$ to connect with others from $Leeds$. Likewise, $\Lambda_{Google}^{Ikea}$ can describe how likely are workers from $Google$ to create links with others from $Ikea$. In sum, the information concerning all homophily factors of a network can be expressed through a group-openness matrix $\boldsymbol{\Lambda_{7x7}}$ as shown in Fig~\ref{fig:matrix}. Consequently, the total homophily factor between a node $i$ and another node $j$ can be defined as:

\begin{ceqn}
\begin{align} \label{eq2}
\mathcal{H}{_P^Q} =& \prod_{\substack{p~\in~P \\ q~\in~Q}} \bm{\Lambda}_{p,q} ~~~~ \parbox{4.5cm}{\footnotesize $P$= groups to which node $i$ belongs\\ $Q$= groups to which node $j$ belongs}
\end{align}
\end{ceqn}

where $P$ and $Q$ are the group sets to which $i$ and $j$ belong, respectively.

\section{Preferential Attachment Model} \label{pa_model}

In the Barabasi and Albert model, a node's probability of creating new connections with others depends exclusively on its degree. However, Eq. \ref{eq2} can be used to introduce the role of attribute similarity in the estimation of the attachment probability between network members. Consequently, the probability $\Pi_i^{PQ}$ that a new node of group-set $Q$ is linked to node $i$ of group-set $P$ is defined as:

\begin{ceqn}
\begin{align} \label{eq3}
\Pi_{i}^{PQ}=& \frac{k_{i}^{P}.\mathcal{H}{_P^Q}}{\sum_{j} k_{j}^{M}.\mathcal{H}{_M^Q}} && \parbox{4.1cm}{\footnotesize $k_i^P$= degree of node $i$ from\\ group-set $P$ \\ $\mathcal{H}{_P^Q}$= total homophily factor be-\\tween group-set $P$ and group-set $Q$}
\end{align}
\end{ceqn}

where $k_i^P$ is the degree of node $i$ from group-set $P$, and $\mathcal{H}{_P^Q}$ represents the total homophily factor between the group-set of the new node $Q$ and the group-set of node $i$. This attachment rule describes the case in which a node without any connections is incorporated to the network. Nevertheless, new links may also emerge between existing network members over time. Particularly, such a probability $\Pi_{ij}^{PQ}$ that an existing node $i$ links to another existing one $j$ is defined as: 

\begin{ceqn}
\begin{align} \label{eq4}
\Pi_{ij}^{PQ}=& \frac{k_{i}^{P}.k_{j}^{Q}.\mathcal{H}{_P^Q}}{\sum_{l} \sum_{m>l}k_{l}^{M}.k_{m}^{N}.\mathcal{H}{_M^N}}
\end{align}
\end{ceqn}

where $k_i^P$ and $k_j^Q$ are the degrees of node $i$ and of node $j$ respectively and $\mathcal{H}{_P^Q}$ is the total homophily factor between their corresponding group-sets $P$ and $Q$. This attachment rule together with the one of Eq. \ref{eq3} were used to generate the network topologies analysed in this paper.

\end{appendices}

\end{document}